\documentstyle{article}
\begin{document}
\begin{center}
{\bf {\large
Research News --- Meson scattering at high precision}}

\bigskip

B. Ananthanarayan 

\medskip
Centre for High Energy Physics,\\
Indian Institute of Science,\\
Bangalore 560 012

\bigskip
\end{center}

\begin{abstract}
A fascinating new generation of experiments has determined certain
meson scattering parameters at high precision.  A confluence of
highly sophisticated theory as well as new experimental ideas have
led to this state of affairs, which sheds important light on the
properties of the strong interactions.  A brief review of the
experiments and the theory is presented.
\end{abstract}

\bigskip
The NA48 collaboration in CERN has recently provided information
through two independent measurements of combinations of so-called scattering
lengths associated with the scattering of two pions.  The 
experiment is based on in CERN's highest-intensity proton beamline 
and uses a large and sophisticated detector.  

The first measurement by the NA48/2 experiment~\cite{hep-ex/0511056}
comes from a recently proposed idea due to 
Cabibbo~\cite{hep-ph/0405001} to measure
a `cusp' in the invariant mass distribution of pions resulting
from the decay of the kaons.  
A cusp at an energy corresponding to $2 m_{\pi^+}$ in
the number distribution of the neutral pion pair as a function
of their invariant mass, manifests itself as an abrupt
change in the derivative of the number distribution.  It is a
very fine effect and can be seen only if the sample size of
events that are analyzed is very large.
The event sample here is enormous
and is based on about 27 million `events' 
$K^\pm\to \pi^\pm \pi^0 \pi^0$.  In units in which the mass
of the charged pion $m_{\pi^+}$ is set to unity, they obtain
for the combination of scattering lengths $|a^0_0-a^2_0|$ the
value $0.264\pm 0.015$, an accurate measurement of what was
a rather poorly measured experimental quantity. 

In the expression above, the $a^I_0,\, I=0,2$ are called
scattering lengths, the subscript denoted by $I$ stands for
`iso-spin' and the subscript $0$ denotes the fact that
this is the scattering length of the angular momentum $l=0$
channel (S-wave scattering length).
Pions, which come in three varieties $\pi^+,\, \pi^-,$ and $\pi^0$,
are the lightest of all strongly interacting particles, and are
bound states of quark and anti-quark pairs of u- and d- varieties.
Heavier counterparts where one of these is replaced by the
s- quark are the kaons.  Pions hold the key to our understanding of
the strong interactions, which are resisted analytic solution
in the low-energy domain.  The strong interactions at the microscopic
level are described by quantum chromodynamics (QCD) which is a theory
in which the degrees of freedom are the quarks and gluons, while
at macroscopic length scales one observes mesons (pions, kaons, etc.)
and baryons (protons, neutrons, hyperons, etc.).  
The lightness of the pions on the hadronic scale ($m_{\pi^0}\simeq
135$ MeV, $m_{\pi^+}\simeq 139$ MeV (we have set the velocity of
light, $c=1$, a common convention)) is today understood in terms
of a phenomenon called spontaneous symmetry breaking of a global
symmetry.  It is common in scattering experiments to decompose
the scattering amplitude, {\it viz.}, namely the mathematical expression
that governs the strength with with a projectile is scattered into
a specific direction by a target, into `partial wave amplitudes'
each of which has a definite angular momentum $l\hbar$; the S- wave
corresponds to the part that is independent of the scattering angle. 
Finally, we note here that the pions
lie in an `iso-triplet', correspond to isospin 1.  Therefore
pion-pion scattering amplitudes could carry isospin of 0, 1, 2
(addition of two isospins $I_1$ and $I_2$, imply that the
total isospin could lie between $I_1+I_2$ and $|I_1-I_2|$).
The advantage of the iso-spin amplitudes
is that the amplitudes for all the physical processes, 
$\pi^+ \pi^+ \to \pi^+ \pi^+$,
$\pi^- \pi^+ \to \pi^- \pi^+$,
$\pi^- \pi^- \to \pi^- \pi^-$,
$\pi^+ \pi^- \leftrightarrow \pi^0 \pi^0$, and
$\pi^0 \pi^0 \to \pi^0 \pi^0$ may all be expressed
in terms of these, when the mass difference of the charged
and neutral pions is neglected.

The second technique employed by the NA48 collaboration, results from
which are still preliminary, comes from a rare decay of kaons
in which a lepton and anti-lepton pair and two pions are
produced in the decay.  The rescattering of
the pions in the final state can actually be observed and provides
a sensitive laboratory for the strength of the interaction of
these particles.  The names of the famous scientists
A. Pais and S. Treiman, and those of N. Cabibbo and
A. Maksymowicz are associated with this effect.
Based on this technique and on the analysis of 370, 000
decays a prelimary number for the scattering length
$a^0_0$ is given as $0.256\pm 0.011$, according to summary
talks posted on the web-site of the collaboration in September 2006.
The E865 experiment at the Brookhaven National Laboratory, 
USA also uses the rare kaon decay to analyze 400, 000
events and measured this quantity to be $0.216\pm 0.015$~\cite{hep-ex/0301040}.
These two experiments in addition to low-energy phase shift
also rely on what is known as Roy equation analysis which
is described in some detail later in this article.

Another important experiment called DIRAC which stands for the
Di-Meson Relativistic Atom complex is an experiment that uses highly
sophisticated experimental techniques to get two charged pions
to bind through the electromagnetic interaction to form a so-called
pionium atom which then scatters into two neutral pions in the
ground state of the atom, upon which the electromagnetic interaction
is switched off and the neutral pions then scatter off.
The lifetime of this state provides an accurate measurement
of the same difference of two scattering lengths                 
as in the cusp experiment of NA48,
an effect predicted in a different setting
over 50 years ago by a highly distinguished set of authors:
S. Deser, M. L. Goldberger, K. Baumann and W. E. Thirring~\cite{Deser}.
Based on a harvest of 6, 600 pionium atoms the experiment reports
the value of $0.264^{+0.033}_{-0.020}$~\cite{hep-ex/0504044},  
from a measurement of the lifetime of the ground state of $\sim 2$ fs.
More data from the experiment is presently being analyzed
to bring down the uncertainties.

Pion-pion scattering has long occupied the attention of theorists even
before the advent of QCD.  The reason for this was that it provided
a paradise for theoreticians due to the simplicity of the process,
and the possibility of deploying many powerful theoretical constraints
that follow from general principles.  Notable amongst these was
the application of dispersion relations, relations which follow
from the principle of causality in field theory. 
Loosely speaking dispersion relations arise from the application
of Cauchy's theory of complex variable theory to scattering amplitudes,
when the latter are considered as complex functions of complex energy
arguments.  Other principles go under the names of `crossing symmetry'  
and unitarity.  In the context of pion-pion scattering a system of dispersion
relations were established that entailed the presence of
certain unknown functions of the momentum transfer which limited
the power of the dispersion relations.
In an extraordinary feat in 1971, S. M. Roy used all the general
properties of scattering amplitudes to eliminate all
these problems, and gave a representation that required
the knowledge of the two scattering lengths only, in addition
to the knowledge only of the imaginary parts of the partial waves~\cite{smroy}. 
This further led to a system of coupled integral equations
for all the partial waves of pion-pion scattering.
However, partial knowledge of the low-lying waves and some
theoretical models of the higher waves could be used to
produce a determination of pion scattering lengths.
This program to pin down pion scattering phase shifts
came to be known as Roy equation analysis.
The analysis of
phase shift information provided by the rare $K_{l4}$ decay
and Roy equation analysis
was used to pin down $a^0_0$ to the range $0.26\pm 0.05$             
based on 30, 000 events from the Geneval-Saclay 
experiment~\cite{Rosselet:1976pu},
when activity in the field stopped for a couple of decades.

After the advent of QCD and the subsequent development of
low energy effective theories for pion-pion scattering
there was a resurgence of interest in the subject.
These effective theories exploit the symmetry properties
of the strong interactions to provide a consistent framework
as an expansion in powers of momenta and the quark masses
has come to be known as chiral perturbation theory,
and is identified with the work of
J. Gasser and H. Leutwyler~\cite{Gasser:1984gg}.
At leading order in the low-energy expansion, S. Weinberg gave a
prediction for $a^0_0$ of 0.16~\cite{Weinberg}, while at next to leading order
the number was revised to $0.20\pm 0.01$, which resulted from
the comprehensive analysis by Gasser and Leutwyler.  The presence
of light particles in the spectrum was the culprit for
this substatial revision.  The revision was found to
stay stable at next to next to leading order, for a 
thorough discussion, see, e.g. ref~\cite{Colangelo:2001df}.  Debates
were sparked on what exactly was nature of the QCD ground
state which would protect this prediction; deviations from
conventionally accepted picture of spontaneous symmetry
breaking due to which pions themselves arise as near
massless states were proposed.  Such scenarios would
have predicted higher values of $a^0_0$~\cite{Stern:1995fw}.  
We note here that a 
comprehensive Roy equation analysis tailored to meet
the needs of modern effective field theories 
was recently presented, see ref.~\cite{ACGL}.

These dispersion relations are sufficiently general to
permit an extension into the complex energy plane.  
Typically, the existence 
of singularities known as poles on the second Riemann
sheet represents the formation of bound
states of quarks and anti-quarks, and imply the formation
of an intermediate unstable particle.  From the real
and imaginary parts of the pole position one may deduce
the mass of this new particle and its `width' which is related
to the inverse lifetime of the particle.  
Recently, using the properties above, along with the accurately known solutions
of the Roy equations the pole position of the state
known as $\sigma$ has been determined to high accuracy
in the pion-pion $I=0$ channel.  Note that this 
is a model independent way of establishing 
of what is the lowest-lying state in the
strong interaction spectrum~\cite{Caprini:2005zr}.  The mass
and width are respectively given as $M_\sigma=441^{+16}_{-6}$ MeV and
$\Gamma_\sigma=544^{+25}_{-18}$ MeV.  Special attention may be paid to
the small uncertainties.  

An extension of the study to a more complicated process
that involves the scattering of pions and kaons has also been
carried but due to paucity of experimental data is yet
to see the spectacular success of pion-pion scattering.
Here the comparison of dispersion relations and chiral perturbation theory 
has been performed in ref.~\cite{hep-ph/0012023}. 
A modern Roy-Steiner analysis, the analog of pion-pion
scattering Roy equation analysis was recently
provided in ref.~\cite{hep-ph/0310283}.  On the experimental side, there is
a proposal to produce $\pi K$ atoms, while there is possibility
of measurement of certain phase shift information at the COMPASS
experiment.  Much work remains to be done.
However, the position of the $\kappa$- resonance has also been recently
determined~\cite{hep-ph/0607133}, using the principles analogous
to those that were described in the pion-pion case earlier, using
the accurately known solutions to the Roy-Steiner system.

There are studies on the lattice of scattering lengths and these
have been reviewed in ref.~\cite{hep-ph/0612112} and is beyond 
the scope of the present article.

To summarize, we have pointed out the results emanating from a
series of beautiful new experiments, and shown the confluence of
theory and experiment.  The theory uses effective field theories,
dispersion relations, and together they make possible precise
predictions in a domain long considered too inhospitable for
such a state of affairs.  

\bigskip

{\bf Acknowledgements:}  I thank M. Passera and
S. Vaidya for
a careful reading of the manuscript and comments.  This work is
supported by the Council of Scientific and Industrial Research
and by the Department of Science and Technology, Government of India.

\end{document}